# Passive control of the onset of vortex shedding in flow past a circular cylinder using Slit

Alok Mishra, Mohd. Hanzla, Ashoke De[a)]

*Department of Aerospace Engineering, Indian Institute of Technology Kanpur, Kanpur, 208016, India.*

The present study investigates the passive flow control phenomena over a two-dimensional circular cylinder using numerical simulations in the laminar regime. The aim is to explore one of the passive control techniques, which involves the introduction of a slit to the geometry of the cylinder. The two parameters, slit width ratio S/D (slit width/diameter) and slit angle (measured with respect to the incoming flow direction), play an essential role in determining the trend of critical Reynolds number ($Re_c$). Most of the analysis invokes flow visualization and saturation amplitude methods to obtain the critical Reynolds number (indicative of the onset of vortex shedding) for different cases. Further, Hopf bifurcation analysis using Stuart-Landau equation and global stability analysis confirm the accuracy and consistency in the predicted solutions. The additional amount of flow through slit increases the pressure downstream of the cylinder, which consequently leads to an increase in $Re_c$ of the modified cylinder. The critical Reynolds number increases with S/D of the modified cylinder at 0° slit angle (as an additional amount of flow grows with S/D). The critical Reynolds number shows an increasing trend with the slit angle in the range of S/D= 0.05-0.15 as the fluctuation intensity reduces with slit angle in this range. For S/D=0.15-0.25, the extra amount of flow through slit induces the instability in wake which causes a decrease in $Re_c$ with slit angle. A correlation is obtained that estimates the critical Reynolds number for a given S/D and slit angle.

## I. INTRODUCTION

Flow past a circular cylinder is a simple and classical problem of fluid mechanics, yet provides a fundamental understanding of flow around bluff bodies. Flow characteristics over a circular cylinder have gained significant research interests due to wealthy flow physics including vortex shedding and vortex-induced vibration (VIV), flow instabilities, etc. The study of Roshko[1] provides a helpful and comprehensive review of flow past bluff bodies and circular cylinders as well.

The characteristic of the fluid flow over the bluff body is described by a non-dimensional parameter, Reynolds number (Re) in the absence of any external forces. Flow regimes, which vary from laminar (Re=1) to fully turbulent (Re=$10^7$), exhibit very distinct flow characteristics. At very low Re (<6), flow is attached to the cylinder while it remains two dimensional and steady, thereby exhibiting a single separation point at the rear stagnation point. Further increase in Re (6<Re<47), the flow separates from the body and forms the recirculation bubble (vortex pair) in the downstream of the bluff body, but the flow remains steady and laminar.

___

[a)] Author to whom correspondence should be addressed. Electronic mail: ashoke@iitk.ac.in

As Re increases, beyond a specific value of Re, flow demonstrates a time-dependency with the periodic shedding of the vortices in the downstream of the body. The steady flow transitions to an unsteady beyond a certain Reynolds number. The Re at which, this bifurcation phenomenon takes place is named as critical Reynolds number ($Re_c$).[2,3] For circular cylinder, this bifurcation takes place at Re=47, and the flow remains two dimensional up to Re=188.5.[4] Therefore, in the range of 0<Re<190, there exits two bifurcation points($Re_c$): one at $Re_c$=47 beyond which the vortex shedding starts, but the wake remains two dimensional and at $Re_c$=188.5 beyond which the wake becomes three dimensional. This periodic shedding is the main reason behind the vortex-induced vibration (VIV). The VIV can lead to the wind-induced vibration and fatigue failure of the bluff body structures.[5,6] The Tacoma Narrows bridge disaster and the liquid sodium leakage accident in the fast breeder reactor, MONJU, are some of the examples of fatigue failure due to VIV.[7]

The flows around a circular cylinder with many modifications have been a perennial research topic concentrating on decreasing the magnitude of unsteady forces and VIV. Passive flow control is of interest in this work, where the primary focus is to find the critical Re for the onset of vortex shedding. The flow control over a cylinder can be divided into two categories, namely active control and passive control. **(a)** In the active flow control technique, external energy is provided to the system. Examples of such flow control techniques include suction control,[8] synthetic jet,[9] plasma actuator,[10,] and rotating cylinders.[11] **(b)** The passive control can be achieved by modifying the geometry of the body. Surface modification with roughness, dimple, helical wire, longitudinal groove, splitter plate, slit, segmented and wavy trailing edge, and small secondary control cylinder[12-23] are classic examples of passive flow control methods.

In active control, the primary aim is to control the boundary layer to withstand certain adverse pressure gradient, and thereby controlling/suppressing the vortex shedding (wake characteristics) behind the bluff body. On the other hand, in passive control, geometric modifications are made to control vortex shedding and suppress VIVs. Usually, active flow control techniques appear to be more effective, but it is not cost-effective. For example, the application of passive flow control on an offshore structure, cable-stayed bridge, and a high-rise building is more realistic as compared to the active flow control, which is very complex in nature. The advantage associated with any passive flow control techniques lies in its implementation as well as maintenance without involving extra cost and energy. Since the focus of our work is to investigate the onset of vortex shedding in a slit cylinder, our subsequent discussion is restricted to the slit cylinder only.

A slit cylinder is one of the ways to suppress/weaken the VIV by modifying the critical Re for the onset of vortex shedding. The pioneering investigation of Igarshi[24,25] was motivated to characterize the flow around a circular cylinder with a diametrical slit at higher Re. The range of Re in his experiments was 13,800 to 52,000 for two slit width ratios of S/D=0.080 and 0.185. The experiments revealed three different flow patterns based on the inclination of the slit. Between 0 to 40°, the incoming flow through the slit acts as a self-injecting jet into the wake of the cylinder, and this extra-amount of mass injection increases the pressure at the rear side as compared to the normal cylinder. While the effect of the slit is negligible for 40 to 60°, Igarshi[24-25] observed periodic boundary layer suction and blowing phenomenon with the period of vortex shedding for 60 to 90°. At 45°, the value of time-averaged drag coefficient ($C_D$) is nearly equal to the baseline cylinder case, and the drag ($C_D$) continues to increase with the slit angle beyond this inclination value. Similar observations have also been reported in one of the recent experimental study by Gao *et at.*[26] for the Re = 26,700 at S/D = 0.75.

The numerical work of Baek and Karniadakis[27,] investigated the effect of slit width variation on vortex-induced vibration (VIV) and investigated the modified cylinder at Re = 500 and 1000 for S/D =0.05 to 0.30 at slit angle 0°. This study was reported for low Re, but the considered Re exhibits unsteady wake characteristics. Also, the results illustrate that the slit parallel to incoming flow is very effective in suppressing VIV when compared to a normal cylinder. The vortex shedding pattern of the cylinder with slit is the same as the unmodified cylinder up to S/D = 0.16. However, the wake flow of the cylinder acquires more complex and irregular vortex shedding patterns beyond S/D = 0.16. Another experimental study of Gao *et at.*[28] reported the variation of the slit width for the modified cylinder at higher Re. Their findings also corroborate the observation of Baek and Karniadakis[27] for the reduction of VIV in the slit cylinder as compared to an unmodified cylinder.

The available literature strongly suggests that the slit through the cylinder as a passive control suppresses the vortex-induced vibration and appears to be an effective passive flow control technique. However, none of these investigations discusses the onset of vortex shedding/control of $Re_c$ using slit cylinder or passive techniques. Thus, the onset of vortex shedding requires detailed investigation and has motivated authors to investigate the effect of passive control using slit through the cylinder on the critical Reynolds number. Hence, the current study addresses these questions to the flow past a slit cylinder at low Re. (i) How do the slit ratio and slit angle affect the critical Re of onset of vortex shedding? (ii) Qualitatively, what is the flow characteristics that affect the onset of vortex shedding? (iii) Finally, the analytical correlation estimates the critical Reynolds number for a given S/D and slit angle.

The present study is entirely numerical to investigate the onset of vortex shedding in a slit cylinder for varying S/D ratio and slit angle. In addition, other analysis tools such as saturation amplitude analysis (SAA), Hopf bifurcation analysis using Landau equation, and Stability analysis are utilized to assess the accuracy and sensitivity of the numerical predictions. First, the critical Re (onset of vortex shedding) for flow past a stationary unmodified circular cylinder is obtained. Then, the critical Re for the onset of vortex shedding for varying slit ratio (S/D) and slit angle is investigated. To better understand the dynamics, the detailed flow characteristics are analyzed. Finally, the analytical correlation is obtained that estimates the critical Re for varying S/D ratio and slit angle.

The paper is organized as follows: Section II describes the numerical details along with problem definition, including the governing equations, and the techniques used to solve them. Section III reports the detailed results, followed by the conclusion in section IV.

## II. SIMULATION DETAILS

### A. Governing equations

The non-dimensional governing equations for viscous Newtonian incompressible flows are given by as,

Continuity equation: $\nabla \cdot \bar{u} = 0$ (2.1)

Momentum equation: $\dfrac{\partial \bar{u}}{\partial t} + \bar{u} \cdot \nabla \bar{u} = -\nabla p + Re^{-1} \nabla^2 \bar{u}$ (2.2)

### B. Computational domain and Boundary condition

The rectangular domain of size 100D×100D is utilized in present simulations, which is shown in Figure 1. The origin of the coordinate system is located at the center of the cylinder. The flow inlet and outlet boundaries are located at 50D upstream and downstream from the center of the cylinder. The top and bottom boundaries are also situated at a distance of 50D from the origin.

A uniform fixed value is specified at the inlet for velocity, while the zero-gradient is applied to pressure. At the outlet, the convective boundary condition is applied. Top and bottom boundaries are treated as slip walls, and No-slip boundary condition is enforced at the cylinder.

### C. Numerical details

The governing equations of fluid flow are solved using open source CFD package OpenFOAM[29]. The standard incompressible solver, pisoFoam, is utilized for pressure-velocity coupling. Second-order discretization schemes are used for discretizing all spatial and temporal terms in the governing equation. Two-dimensional computational grids used in the current simulations are generated using ANSYS® ICEM-CFD[30], as shown in Figure 2.

Initially, a grid independence test is carried out using four different grids for Re = 100 for unmodified cylinder or baseline cylinder or normal cylinder, which is shown in Table I. The grids are consistently refined by increasing the nodes around the cylinder surface. In this process, the y + < 1 is maintained for all the grids. Grid independence test is assessed through the time-averaged drag ($C_D$) and the r.m.s value of lift coefficient ($C_{Lrms}$). As observed in Figure 3, there is no significant difference in the results of Grid-3 and 4. Hence, it is concluded that the solutions become grid-independent and converge at Grid-3. Further, Table II reports the comparison of aerodynamic parameters with the existing literature. Notably, the present simulated results provide excellent predictions of the aerodynamic parameters, similar to the predictions reported in the published literature. Hence, grid-3 is selected for further analysis.

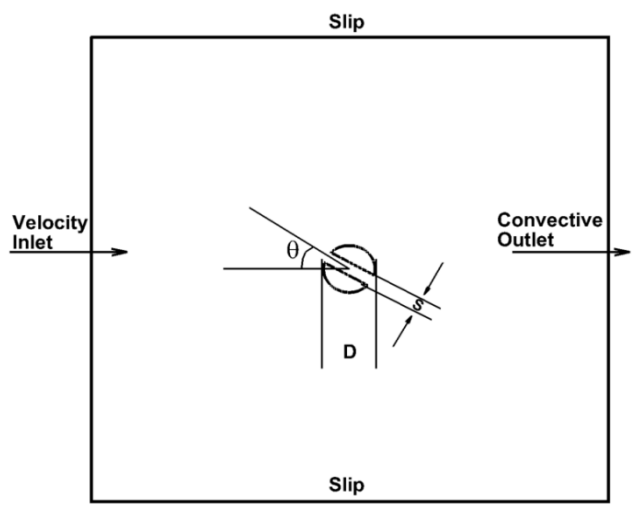

FIG. 1. Schematics of a computation domain with boundary conditions.

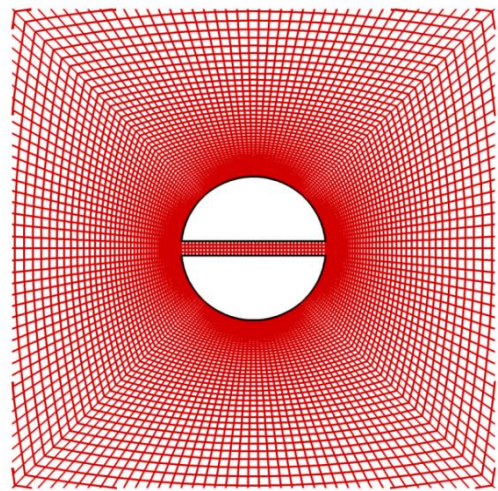

FIG. 2. Computational grid over the cylinder

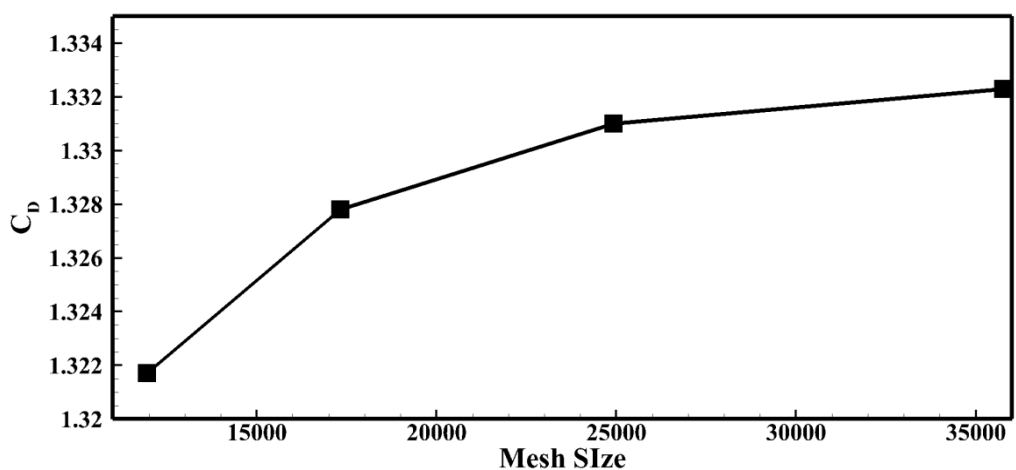

FIG. 3. Grid independence test for baseline cylinder

TABLE I. Grids used for the simulations

| Grid | Cell No. | $C_D$ | % change ($C_D$) | $C_{Lrms}$ | % change ($C_{Lrms}$) |
|---|---|---|---|---|---|
| Grid-1 | 11956 | 1.3217 | - | 0.2280 | - |
| Grid-2 | 17324 | 1.3278 | 0.46 | 0.2326 | 2.01 |
| **Grid-3** | **24920** | **1.3310** | **0.24** | **0.2357** | **1.33** |
| Grid-4 | 35760 | 1.3323 | 0.09 | 0.2376 | 0.806 |

TABLE II. Comparison of baseline cylinder results with different study

| **Available works of literature** | $C_D$ | $C_{Lrms}$ | **Strouhal number (St)** |
|---|---|---|---|
| Park et al.[31] | 1.33 | 0.235 | 0.165 |
| Mittal[32] | 1.322 | 0.2451 | 0.164 |
| Stalberg et al.[33] | 1.32 | 0.233 | 0.166 |
| Qu et al.[34] | 1.319 | 0.225 | 0.164 |
| Posdziech and Grundmann[35] | 1.325 | 0.228 | 1.64 |
| **Present (Mesh 3)** | **1.3310** | **0.2357** | **0.1642** |

A Grid convergence study is also performed using Grid Convergence Index (GCI) as proposed by Roache[36]. Richardson extrapolation is used to evaluate grid refinement study for different grid solutions with uniform grid spacing increments. Grid-3 (medium grid) is chosen as a base grid for the GCI. The Richardson error estimator for Grid-3 and Grid-4 (fine grid) is expressed as,

$$E^{fine} = \frac{\varepsilon_{43}}{1-r^o} \qquad (2.3)$$

The coarse grid Richardson error estimator for Grid-2 (coarse grid) and base grid (Grid-3) is described as,

$$E^{coarse} = \frac{r^o \varepsilon_{32}}{1-r^o} \qquad (2.4)$$

Where, the error ($\varepsilon$) is determined through the discrete solution (*f*) of two consecutive grids which is represented as follows,

$$\varepsilon_{i+1,i} = \frac{f_{i+1} - f_i}{f_i} \qquad (2.5)$$

The *r* is known as the grid refinement ratio between two successive grids which is specified as,

$$r_{i+1,i} = \frac{h_{i+i}}{h_i} \qquad (2.6)$$

Here, *h* is the grid spacing for a particular grid.

The CGI provides a uniform reporting of grid convergence test for progressive refinement of the grid. The GCI for the fine grid and the coarse grid is provided as Ref.[36],

$$GCI^{fine} = F_s \left| E^{fine} \right| \quad (2.7)$$

$$GCI^{coarse} = F_s \left| E^{coarse} \right| \quad (2.8)$$

Where $F_s$ is known as the factor of safety. The value of $F_s$ is taken equal to 1.25, as suggested by Wilcox[37].

The convergence conditions of the system must be verified before using Richardson extrapolation for GCI. The convergence conditions are shown as,

- Monotonic convergence: $0 < R < 1$
- Oscillatory convergence: $R < 0$
- Divergence: $R > 1$

Where $R$ is convergence ratio which is described as,

$$R = \frac{\varepsilon_{i,i-1}}{\varepsilon_{i+1,i}} \quad (2.9)$$

The order of accuracy of the numerical scheme (º) is evaluated through the $L_2$ norms of the errors between the grids. The $L_2$ norm for the two consecutive grids is defined as,

$$L_2 = \sqrt[2]{\left( \sum_{j=1}^{N} \left| \varepsilon_{i+1,i} \right|^2 / N \right)} \quad (2.10)$$

Where N is the total number of grid points utilized to determine the $L_2$ norm. The results are plotted against the theoretical 2nd order slope, as shown in Figure 4. The order of accuracy for the numerical schemes is found to be 1.93. This order of accuracy used to determine the GCI for different grids. The value of $R$ (in Table III) confirms the monotonic convergence criteria for the grids.

Table III shows the GCI for the time-averaged drag ($C_D$) and the r.m.s value of lift coefficient ($C_{Lrms}$) from three different grids (Grid-2, Grid-3, and Grid-4). The value of GCI reduces for the consecutive grid refinements ($GCI^{fine} < GCI^{coarse}$) for both the parameter $C_D$ and $C_{Lrms}$. It can be concluded that the dependency of the numerical solution on the mesh size is decreased. The large decrement is found in the value of $GCI^{coarse}$ to $GCI^{fine}$; therefore, the grid is nicely converged for Grid-3. The GCI results also corroborate the results of the grid independence test, as shown in Table I.

TABLE III. Richardson error estimation and Grid-Convergence Index for three sets of grids

| | $r_{32}$ | $r_{43}$ | (°) | $\varepsilon_{32}$ (10$^{-2}$) | $\varepsilon_{43}$ (10$^{-2}$) | $R$ | $E^{coarse}$ (10$^{-2}$) | $E^{fine}$ (10$^{-2}$) | $GCI^{coarse}$ (%) | $GCI^{fine}$ (%) |
|---|---|---|---|---|---|---|---|---|---|---|
| $C_D$ | 1.2 | 1.2 | 1.93 | 0.241 | 0.0977 | 0.405 | -0.812 | -0.231 | 1.015 | 0.289 |
| $C_{Lrms}$ | 1.2 | 1.2 | 1.93 | 1.332 | 0.806 | 0.604 | -4.490 | -1.911 | 5.613 | 2.388 |

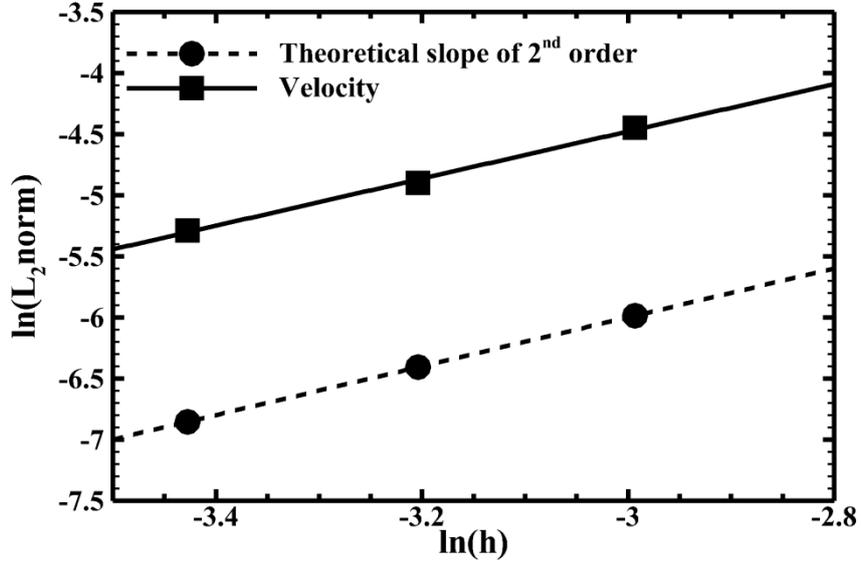

FIG. 4. log-log plot of L$_2$ norm against the grid spacing h

## III. RESULTS AND DISCUSSION

We initially present our results using different techniques such as flow visualization, saturation amplitude analysis (SAA), Hopf bifurcation analysis using Landau equation, and Stability analysis. The flow visualization method provides the range for the critical Reynolds number for a particular case (for example, Re$_c$=47-48 for S/D=0 and θ=0 case) while the saturation amplitude method (SSA) determines the exact value of the Re$_c$ (equals to 47.02 for S/D=0 and θ=0° case). The Hopf bifurcation analysis using Landau equation and Stability analysis are applied to confirm the accuracy and consistency of the simulated results. Later on, the effect of the S/D ratio and slit angle is investigated for the onset of vortex shedding. The slit ratio is varied from 0.05 to 0.25 with the step increment of 0.05 for a fixed slit angle, and for every S/D, the slit angle is varied from 0° to 50° with the step increment of 10°.

### A. Validation and Verification of numerical approach

#### 1. Flow field visualization method (FVM)

A systematic simulation has been carried out with an increment of ΔRe =1 for baseline circular cylinder to evaluate the critical Re. Figure 5 illustrates the flow field for the normal cylinder and flow control with a slit angle at 0°. Figure 5(a) depicts the vorticity contour and streamlines plot for Re = 47, which confirms the symmetric behavior of the flow field at this Reynolds number, and the recirculation length of the bubble happens to be the same behind the cylinder. Whereas, Figure 5(b) exhibits the asymmetric streamlines for Re = 48, where the recirculation length of the bubble is different behind the cylinder. The vorticity contour plots also support the streamline plot results. The vortex shedding at Re = 48 can be seen downstream of the cylinder in vorticity contour plots, and this confirms that flow shows unsteady behavior at this Re for baseline circular cylinder. Based on the observation, the critical Re for a normal cylinder is found to be Re = 47, which is in good agreement with the past works of literature .[2,3]

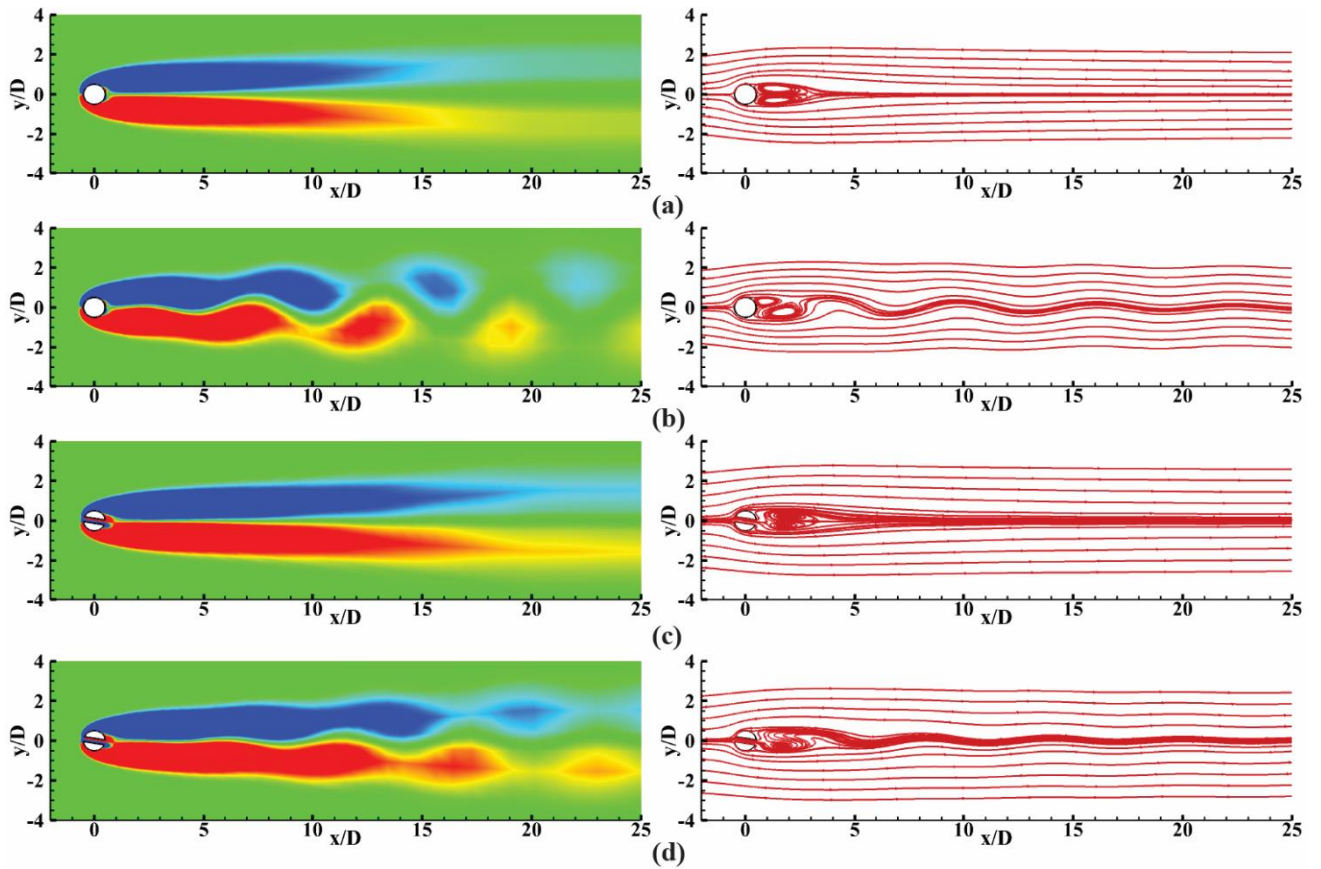

FIG. 5. Instantaneous vorticity contour and streamline pattern of (a) Baseline cylinder Re=47 and (b) Baseline cylinder Re=48 (c) Slit width (S/D=0.25 & θ=10) Re=63 (d) (S/D=0.25 & θ=10) Re=64

The surface of the cylinder has been modified using a slit parallel to the income flow direction, which is known as passive control of flow. The slit along the cylinder behaves like a self-injecting jet into the wake region. This extra amount of energy in the system increases the pressure of flow downstream of the cylinder, which maintains symmetry in flow and delay in the bifurcation. Consequently, there is a significant increase in the critical Reynolds number.

Figure 5 (c-d) shows the streamline and vorticity contour plot for a modified cylinder with a slit width of 0.25 at a slit angle $10^0$. As observed, the flow is symmetric up to Re = 63 and asymmetric at Re = 64. Precisely, the flow bifurcation takes place at Re = 63. This is true for this particular S/D and slit angle. This exercise is carried out for the prediction of $Re_c$ using flow visualization method, and the findings are consistent with other methods adopted herein (Table VII).

**2. Saturation amplitude analysis (SAA)**

One can determine flow stability over the cylinder with this method. In this approach, the instantaneous lift coefficient curve is utilized to examine whether the flow exhibits time dependency or not. When the lift coefficient decays with time, the flow remains stable, as shown in Figure 6 (a). If the flow becomes unstable, then the lift coefficient curve grows with respect to time and attain the saturated value (Figure 6 (b)). This value is known as the saturation amplitude ($\wedge_{sat}$), and the saturation amplitude increases with the Re. It can be inferred that the value of saturation amplitude becomes 0 for the critical Re.[38] The simulation at various Reynolds numbers provides the value of saturation amplitude corresponding to the particular Re. Similarly, the value of saturation amplitude is evaluated for different cases of the slit width and slit angle. Figure 7 presents the saturation amplitude analysis to find out the value of critical Re. The symbol represents the value of the $\wedge_{sat}^2$ for corresponding Re, and the dotted line exhibits the least square curve fitted to the $\wedge_{sat}^2$. The value of critical Re for the normal cylinder is found to be 47.02 (Figure 7(a)), which is in excellent agreement with the flow visualization method and previously published work (Table II). Figure 7(b) provides the saturation amplitude analysis for the S/D = 0.25 at 10° of slit angle. Notably, the value of critical Re is found to be 63.68. The flow visualization method predicts that the value of critical Re is 63, which is also in good agreement with this approach (Table VII). The saturation amplitude analysis is not valid for the S/D = 0.25 at 0° slit angle case. The variation in the value of saturation amplitude is negligible for a whole range of Reynolds number near Hopf bifurcation (from Re = 70 to 78). The least-square fit curve is almost horizontal in this case, and the value of critical Re is incorrect. This is the limitation of the saturation amplitude method. It cannot predict the critical Re for and above S/D = 0.25 properly.

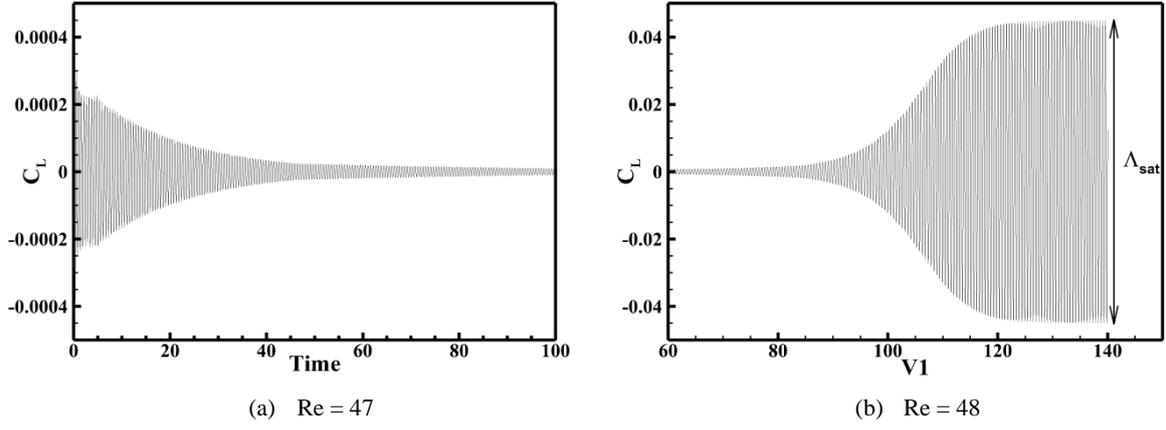

(a) Re = 47

(b) Re = 48

FIG. 6. Lift coefficient time history for different Re: (a) Re=47. (b) Re=48

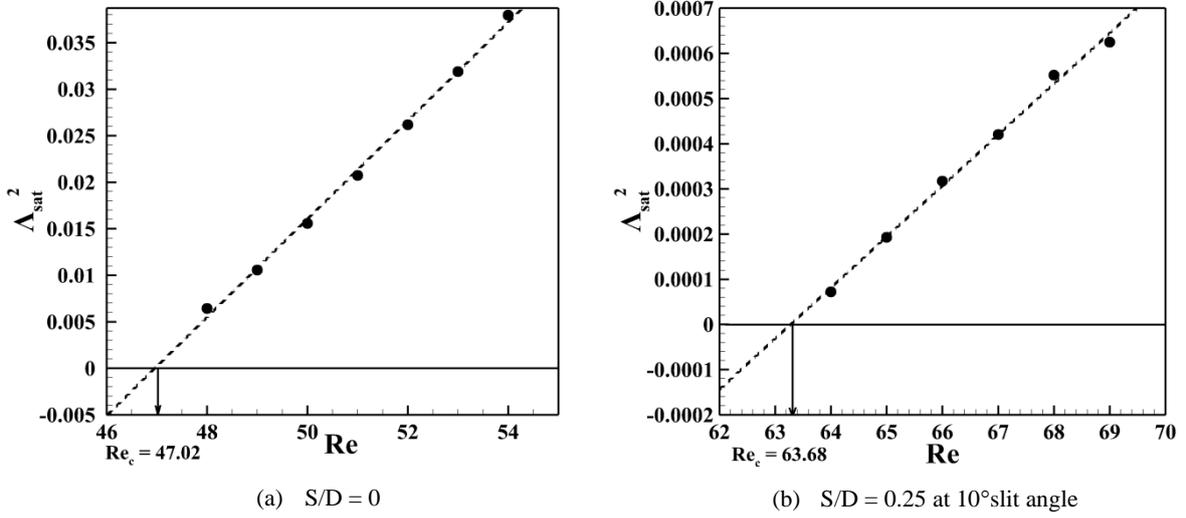

(a) S/D = 0

(b) S/D = 0.25 at 10°slit angle

FIG. 7. Saturation amplitude analysis: (a) S/D=0, (b) S/D=0.25

### 3. Hopf bifurcation analysis (HBA)

The supercritical Hopf bifurcation can be evaluated using the Landau equation. The Landau equation has been applied to analyze the flow stability over the cylinder in previously published works.[38-41] The signal with slowly varying amplitude A(t) is considered as a Landau equation which is written as:

$$\frac{dA}{dt} = \lambda A - c|A|^2 A \tag{3.1}$$

Where, $\lambda$ and c are constants, and describes as $\gamma + i\omega$ and $\alpha + i\beta$, respectively. The amplification rate and the angular frequency of the signal are represented through $\gamma$ and $\omega$. The values of these constants are in global nature and provide the same at all points of the flow field. The other two constants $\alpha$ and $\beta$ have different values at various locations in the flow field. The saturation amplitude given by the Landau equation is calculated using the formula, $\Lambda_{sat} = (\gamma/\alpha)^{1/2}$.

The amplitude variation in the Landau equation can be expressed as follows:

$$A(t) = \Lambda(t)e^{i\theta(t)} \tag{3.2}$$

The substitution of the amplitude in equation (3.1) will provide the pair of equations as follows

$$\frac{d(\ln(\Lambda))}{dt} = \gamma - \alpha\Lambda^2 \tag{3.3}$$

$$\frac{d\theta}{dt} = \omega - \beta\Lambda^2 \tag{3.4}$$

The signal decays with time known as a stable system. The left side of equation (3.3) is known as the instantaneous growth rate, which should be negative; therefore, $\gamma - \alpha\Lambda^2$ must also be negative for the stable system. Therefore, the stable condition is achieved when the amplification rate $\gamma < 0$ and $\alpha > 0$. The system becomes unstable when the value of $\gamma$ is positive, so the amplitude grows with time. It can be concluded that the value of Re, where the amplification rate changes sign, is considered as a critical Re.

Figures 6(a) and 6(b) show the lift coefficient history, which is used as an input signal for the Landau equations. To solve the equations, the local maxima and minima are identified from the lift coefficient signal. This provides the instantaneous values of amplitude and frequency with respect to time. Figures 8(a) and 8(b) show least square fit on the data point for the plot for $d(\ln\Lambda)/dt$ vs $\Lambda^2$ and $d\theta/dt$ vs $\Lambda^2$. The slope and intersection of plots provide the values of constant $\gamma$, $\alpha$, $\omega$ and $\beta$. The constant of saturation ($C_\infty$) is described as the ratio of $\beta/\alpha$ and the variation of angular frequency at saturation ($\Delta\omega$) is determined as $\Delta\omega = -\gamma C_\infty$. The calculated landau constants are in good agreement with published work (Table IV). The variation of amplification rate ($\gamma$) and angular frequency with Reynolds number for the unmodified cylinder is also shown in Table V. As observed, the amplification rate ($\gamma$) increases with the Re while the negligible effect in angular frequency is witnessed with Re. The amplification rate ($\gamma$) is found to be negative for Re = 47, which means flow remains stable at this Re. The flow becomes unstable for Re = 48 due to the positive value of $\gamma$. Figure 9(a) depicts the least square fit for the amplification rate ($\gamma$) for different Re. The critical Reynolds number is obtained where the amplification rate becomes equal to zero, which is obtained 47.05 for the normal cylinder. This model is also applied to the passive flow control cases. Figures 8(a) and (b) also show least square fit on the data point for the plot for $d(\ln\Lambda)/dt$ vs $\Lambda^2$ and $d\theta/dt$ vs $\Lambda^2$ for modified slit case (S/D = 0.05 with $0^0$ slit angle). It is observed from Figure 9(b) that the $Re_c$ is found to be at 47.60, which is in excellent agreement with the results obtained using the SAA method, as shown in Table VII.

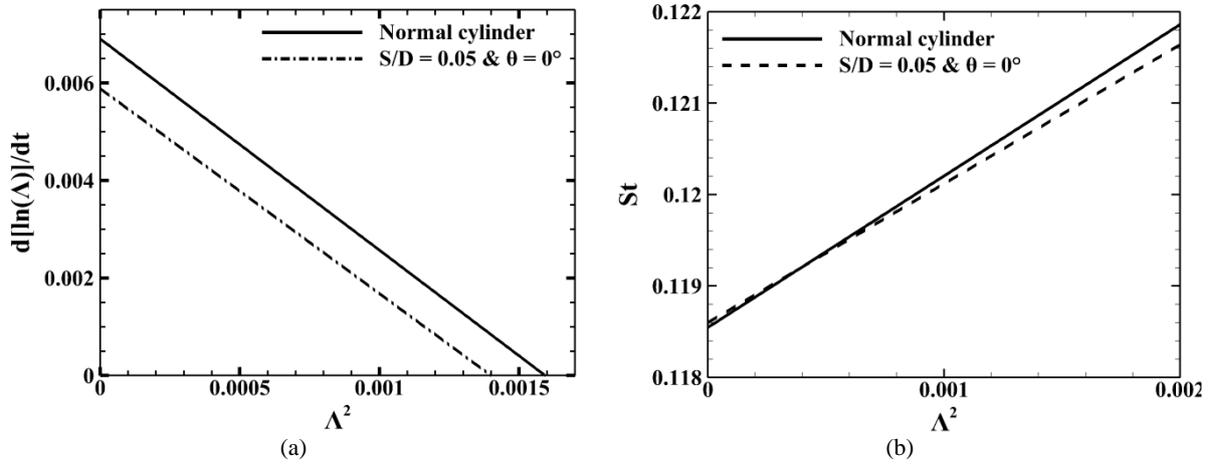

FIG. 8. (a) variation of $d(\ln\Lambda)/dt$ against $\Lambda^2$ and (b) variation of St against $\Lambda^2$ (Re = 48)

TABLE IV. Comparison Landau constant with published literature

| Reference | γ | ω | $C_\infty$ | Δω |
|---|---|---|---|---|
| Kumar and Biswas[41] | 0.11434 | 0.8341 | -1.817 | 0.01967 |
| Paul et al.[38] | 0.007414 | 0.8130 | -2.381 | 0.01874 |
| Dusek et al.[42] | 0.007931 | 0.7411 | -2.709 | 0.02144 |
| **Present** | **0.00691** | **0.7448** | **-2.393** | **0.01653** |

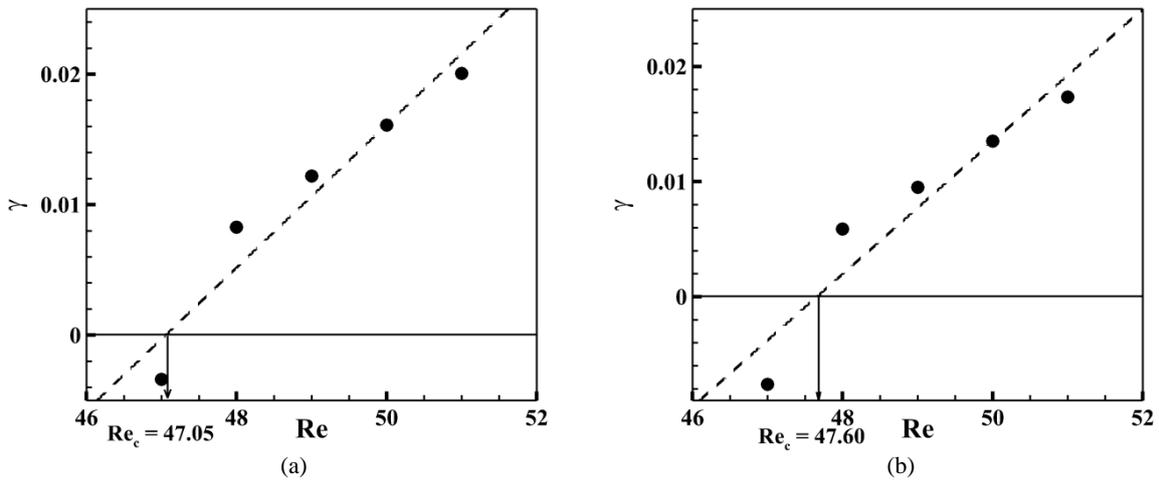

FIG. 9. Calculation of $Re_c$ for (a) S/D=0 (b) S/D=0.05

TABLE V. Variation of amplification rate (γ) and angular frequency (ω) with Re for S/D = 0

| Re | γ | ω |
|---|---|---|
| 47 | -0.0034 | 0.7382 |
| 48 | 0.00827 | 0.7427 |
| 49 | 0.0122 | 0.7451 |
| 50 | 0.0161 | 0.7483 |
| 51 | 0.02005 | 0.7514 |

### 4. Stability analysis (SA) through NEK5000

To confirm and check the accuracy of the predicted results with other methods, the linear stability analysis for the normal cylinder and the modified cylinder is also performed. The advance linear stability tool is used to investigate the flow stability over the cylinder, which is based on the spectral element method code NEK5000. This global linear stability tool is developed by Peplinski *et at.*[43] The linearized Navier-Stokes (LNS) equation is considered for the stability calculation. The steady-state solution of non-linear Navier-Stokes equation is used as a base flow for the calculation. This base flow is computed through selective frequency damping (SFD).[44] The linearized incompressible Navier-Stokes equations are recast as follows,

$$\frac{\partial \vec{u}'}{\partial t} + \vec{u}'.\nabla \overline{U} + \overline{U}.\nabla \vec{u}' - \frac{1}{Re}\nabla^2 \vec{u}' + \nabla p' = f \qquad (3.5)$$

$$\nabla.\vec{u}' = 0 \qquad (3.6)$$

Where, the equations are linearized about base flow $U$ (a function of space only) with $u'$ (velocity perturbation-function of space and time), $p'$ (pressure perturbation), and Re (Reynolds number). $f$ is the forcing term that normally vanishes inside the computational domain but can be utilized as sponge layers at the inflow/outflow boundary.

The LNS can be rewritten as an eigenvalue problem, and the matrix free method adopting the time-stepping Arnoldi approach is used to calculate eigenvalues for the dynamic of nonlinear flows. The eigenvalue evaluates the state of the system (stable or unstable). If one of the eigenvalues is found to be greater than zero, then the instability grows exponentially, and finally, the flow becomes unstable. If the calculated eigenvalues are negative, then the instability dies down, and flow remains stable.

Before moving ahead with the stability calculations, the flow field results on a normal cylinder are verified using the NEK5000. For normal cylinder, the simulation are performed for Re = 47, 48, 49, 50, 51, and 52. Instantaneous vorticity contour plot (Figure 10) shows a stable solution for Re = 47, and the flow becomes unstable for Re = 48. The critical Re for the normal cylinder is found to be at Re = 47, which is also corroborated with the results from the flow visualization method (section III(A-1)). More precisely, the critical Re can be calculated from the saturation amplitude method (SAA), which provides validation of results from NEK5000. The least-square of saturation amplitudes ($\wedge^2_{sat}$) for various Reynolds numbers are plotted in Figure 11. It can be seen that the critical Re for the normal cylinder is 47.11 and, the error in critical Re between FVM and NEK5000 is only 0.19%. It can be concluded that the flow field from NEK5000 exhibits similar to OpenFOAM.

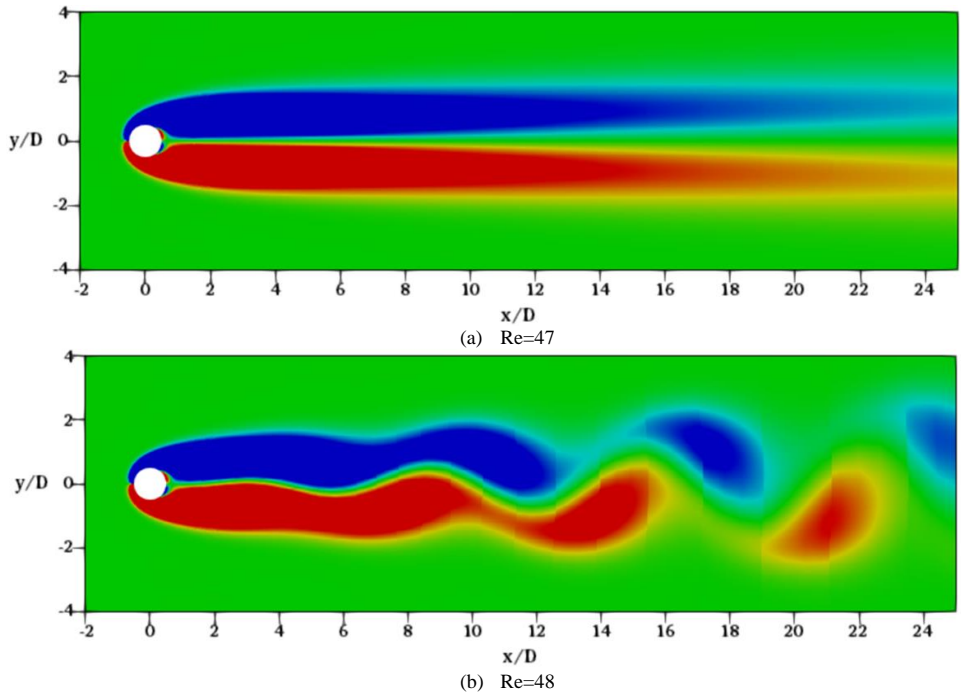

FIG. 10. Instantaneous vorticity contour of Baseline cylinder: (a)Re=47, (b) Re=48

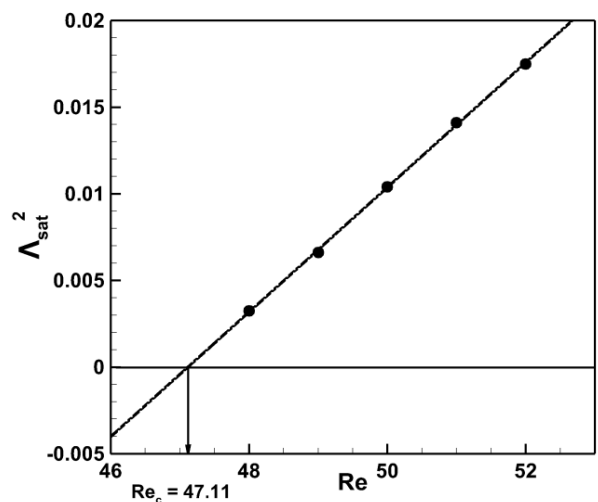

FIG. 11. Saturation amplitude analysis for a normal cylinder of NEK5000 results

The base flow has been evaluated with selective frequency damping (SDF) using the NEK5000 platform. The eigenvalue problem is solved using the implicitly restarted Arnoldi Method with the ARPACK library. Figure 12 and 13 presents the global stability analysis results for modified (S/D=0.20 at $0^0$ slit angle) and normal cylinders. All the eigenvalues are found to be negative for Re = 47, as shown in Figure 12(a); therefore, the system remains in stable condition in this Re. Figure 12(b) shows that one of the calculated eigenvalues is exhibited a positive value for Re = 48. The instability increases exponentially for this Re (=48), and flow becomes unstable. The stability analysis results provide excellent predictions of the flow field, similar to the predictions reported in the published literature, as shown in Table VI. The value of $Re_c$ is found to be 47.04 for the normal cylinder, as tabulated in Table VII. Upon validation, the stability analysis has been successfully extended for a modified cylinder for the S/D = 0.20 with a slit angle of 0° (Figure 13). Interestingly, one can notice that the bifurcation takes place at Re=51.90. The result of the stability analysis corroborates with the flow visualization method, saturation amplitude, and Hopf bifurcation analysis, as shown in Table VII. Therefore, the stability analysis is utilized for selective cases only as it is computationally quite expensive.

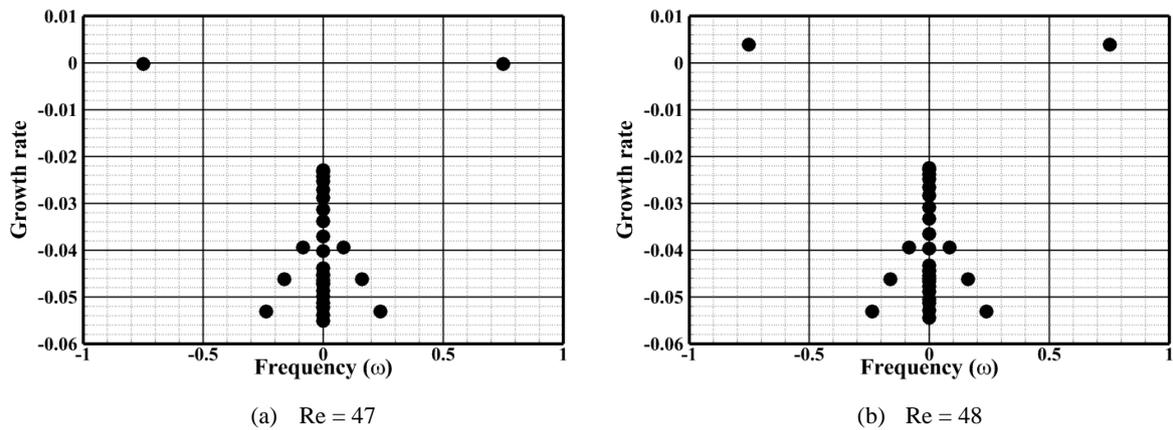

FIG. 12. spectra (growth rate vs. frequency) for different cases of the normal cylinder: (a)Re=47, (b)Re=48

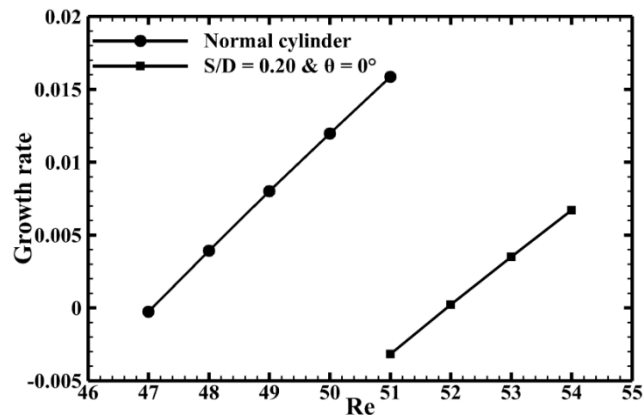

FIG. 13. Growth rate vs. Re for normal cylinder and S/D=0.20 at 0° slit angle

TABLE VI. Comparison of Strouhal number (St) with previous studies

| Investigations | Rotation (ω) | Strouhal number (St) |
|---|---|---|
| Dusek et al[42] | 0.741 | 0.1179 |
| Paul et al.[38] | 0.8130 | 0.1293 |
| Kumar and Biswas[41] | 0.8341 | 0.1327 |
| **NEK5000** | **0.7516** | **0.11962** |
| **HBA** | **0.7448** | **0.11855** |

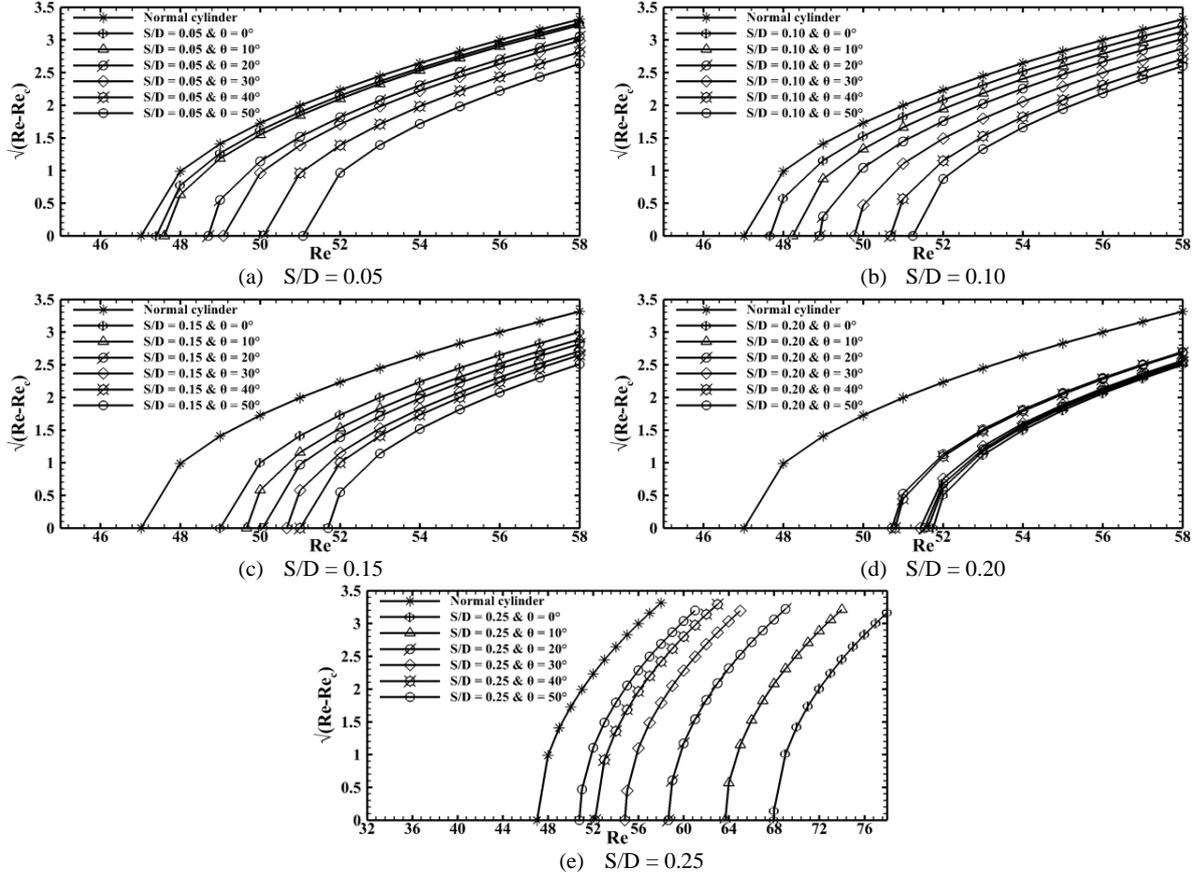

FIG. 14. Bifurcation diagram for normal and modified cylinder for different S/D: (a)0.05, (b)0.10, (c)0.15, (d)0.20, (e) 0.25

## B. Comparative analysis

Here, Table VII shows the results of the flow visualization method, saturation amplitude analysis and Hopf bifurcation analysis for all the combination of slit ratios and slit angles. As mentioned earlier, only a few selected cases are reported for the stability analysis. Upon analyzing these results, the critical Reynolds number is found almost the same for all four different methods. The stability analysis is applied to the normal cylinder and one case of a slit cylinder to verify the bifurcation point, which is evaluated through other methods. The bifurcation diagram of the normal cylinder and modified cylinders are exhibited in Figure 14. As seen, the critical Re grows with a slit angle for S/D = 0.05-0.15 and goes down with a slit angle for S/D = 0.20-0.25. This phenomenon is elaborated in detail in section III(D).

TABLE VII. Comparison of critical Reynolds number evaluated with four different methods

| Slit width S/D | Slit angle (in degrees) | FVM | SAA | HBA | Stability analysis |
|---|---|---|---|---|---|
| 0 | - | 47.5±0.5 | 47.02 | 47.05 | 47.04 |
| 0.05 | 0 | 47.5±0.5 | 47.40 | 47.60 | |
| | 10 | 47.5±0.5 | 47.60 | 47.76 | |
| | 20 | 48.5±0.5 | 48.70 | 48.35 | |
| | 30 | 48.5±0.5 | 49.08 | 48.67 | |
| | 40 | 49.5±0.5 | 50.08 | 49.47 | |
| | 50 | 50.5±0.5 | 51.07 | 50.81 | |
| 0.10 | 0 | 47.5±0.5 | 47.67 | 47.77 | |
| | 10 | 48.5±0.5 | 48.24 | 48.13 | |
| | 20 | 48.5±0.5 | 48.91 | 48.58 | |
| | 30 | 49.5±0.5 | 49.78 | 49.51 | |
| | 40 | 50.5±0.5 | 50.68 | 50.40 | |
| | 50 | 51.5±0.5 | 51.24 | 51.11 | |
| 0.15 | 0 | 49.5±0.5 | 49.00 | 48.62 | |
| | 10 | 49.5±0.5 | 49.66 | 49.45 | |
| | 20 | 50.5±0.5 | 50.06 | 49.82 | |
| | 30 | 50.5±0.5 | 50.67 | 50.30 | |
| | 40 | 51.5±0.5 | 51.00 | 50.82 | |
| | 50 | 51.5±0.5 | 51.70 | 51.45 | |
| 0.20 | 0 | 51.5±0.5 | 51.75 | 51.70 | 51.90 |
| | 10 | 51.5±0.5 | 51.62 | 51.41 | |
| | 20 | 51.5±0.5 | 51.54 | 51.28 | |
| | 30 | 51.5±0.5 | 51.43 | 51.12 | |
| | 40 | 50.5±0.5 | 50.80 | 50.79 | |
| | 50 | 50.5±0.5 | 50.72 | 50.66 | |
| 0.25 | 0 | 68.5±0.5 | NA | 67.98 | |
| | 10 | 63.5±0.5 | 63.68 | 63.40 | |
| | 20 | 58.5±0.5 | 58.63 | 58.20 | |
| | 30 | 54.5±0.5 | 54.80 | 54.24 | |
| | 40 | 51.5±0.5 | 52.15 | 51.60 | |
| | 50 | 50.5±0.5 | 50.78 | 50.41 | |

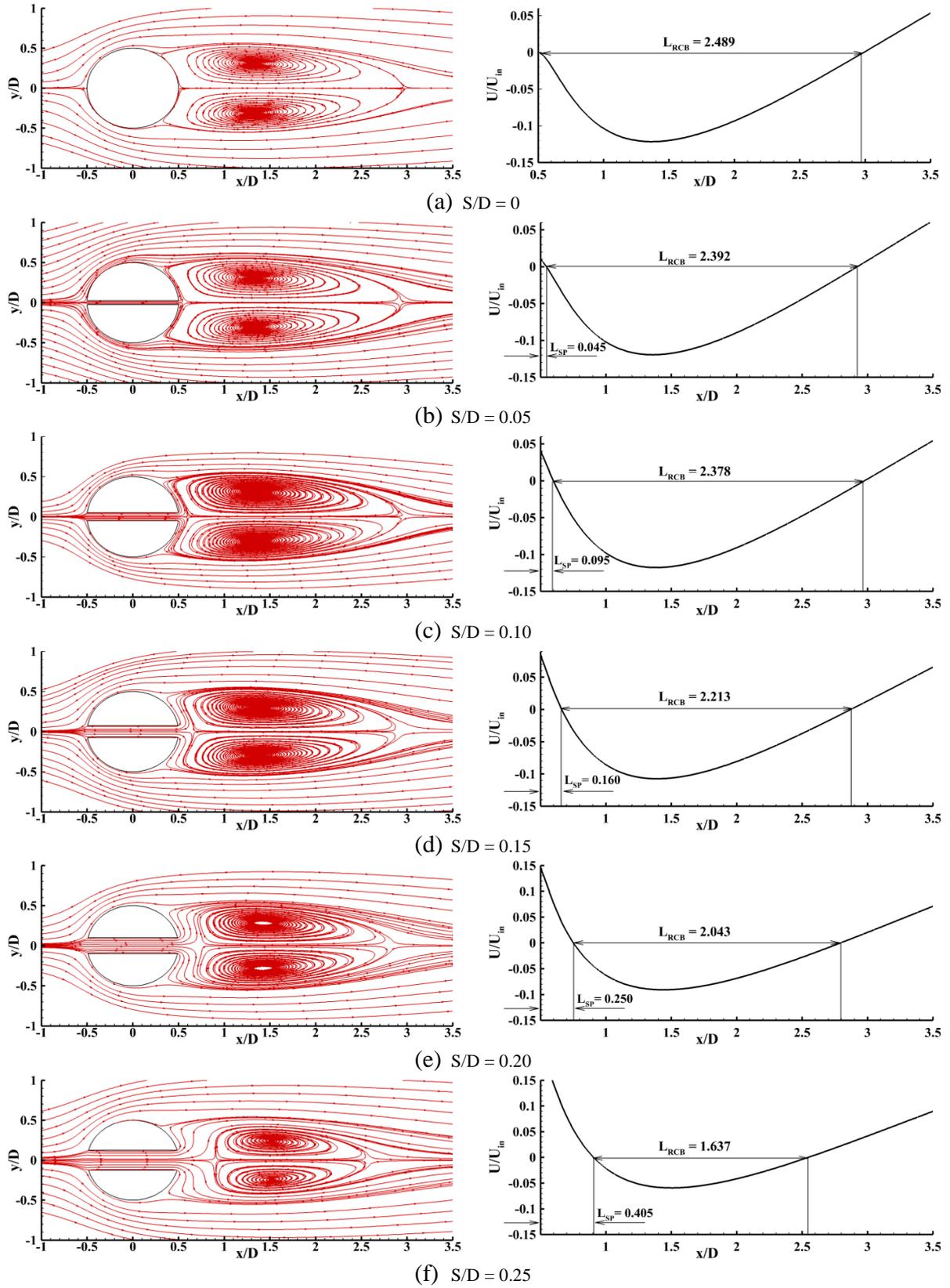

FIG. 15. Streamline pattern for varying S/D ratio at Re = 47: (a) 0.0, (b) 0.05, (c) 0.10, (d) 0.15, (e) 0.20, (f) 0.25

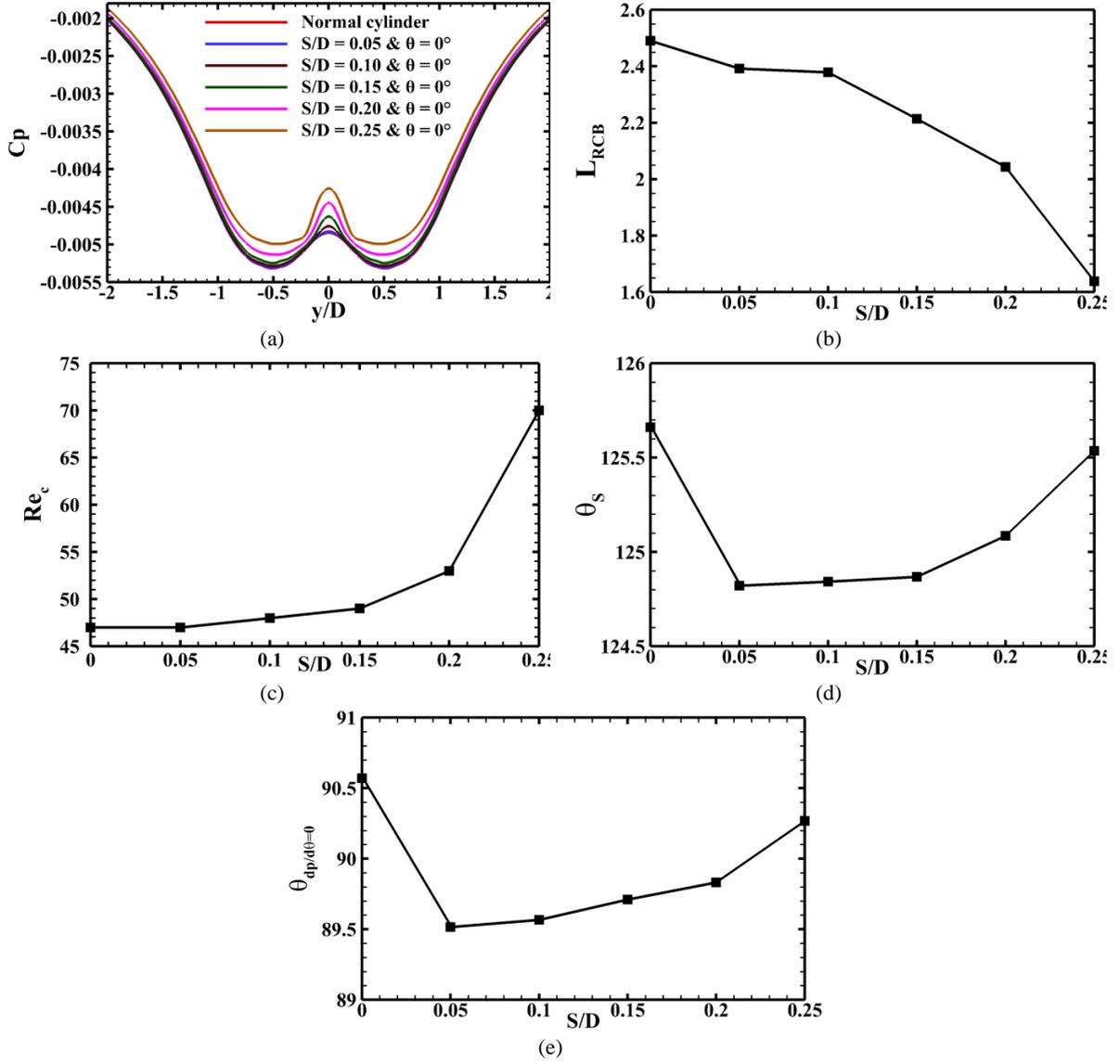

FIG. 16. (a) Coefficient of pressure at downstream of the cylinder for normal and modified cylinder (at x/D=0.6) (b) variation of circulation bubble length with S/D at $\theta=0^0$ (c) Variation of Re$_c$ with S/D at $\theta=0^0$ (d) variation of separation angle on cylinder with S/D at $\theta=0^0$ (e) variation of angle where value of pressure gradient becomes zero with S/D at $\theta=0^0$

## C. Effect of slit width (S/D) variation

The analysis is extended further for five different slit width ratio ranges from 0.05 to 0.25 at Reynolds number 47 to determine the effect of slit width variation on Re$_c$. At this Reynolds number flow remains at steady state for all the cases, so it is the best position to compare the flow physics for all the cases. The effect of the flow control with slit width variation is shown in Figure 15. The extra amount of flow through slit displaces the location of vortex formation downstream, and the saddle point also moves away from the cylinder, as shown in Figure15. This phenomenon can be explained through the coefficient of pressure comparison between the normal cylinder and modified cylinders with various S/D as shown in Figure 16(a). It can be observed that the coefficient of pressure increases with the slit width. The pressure recovery takes place by the extra amount of flow through the slit. Due to pressure recovery at the downstream of the cylinder, the recirculation bubble pushes away from the cylinder. The recovery in pressure also causes a decrease in the recirculation bubble length ($L_{RCB}$) (Figure 16 (b)).

The base suction of the cylinder increases with the Re for the normal cylinder case, which develops instability at the recirculation region from the end of the recirculation bubble. The strength and amplification of the instability grow with Re, which leads to laminar vortex shedding.[45-48] One can derive from the above discussion that base suction is directly related to the critical Re. Thus by controlling the base suction, one can shift the Hopf bifurcation point of the cylinder. The extra energy provided by slit decreases the base suction, which in turn results in reduced pressure drag, as shown in Table VIII. This is the reason behind the increase in the bifurcation point using slit as a passive control over the cylinder. The variation of critical Re with slit width is depicted in Figure 16 (c).

The flow separation angle is also affected by the passive control method. Figure 16 (d) reports the variation of the flow separation angle with slit width. Interestingly, the flow separation angle suddenly decreases by using the slit as passive control on a normal cylinder. The maximum decline is found in the slit width of 0.05 as compared to the normal cylinder case. And the separation angle increase with the slit width. This trend can be explained through the pressure gradient over the cylinder. Figure 16 (e) shows the value of the angle ($\theta_{dp/d\theta=0}$) over the cylinder, where the pressure gradient becomes equal to zero. It can be seen that the variation of $\theta_{dp/d\theta=0}$ with slit width follow a similar trend as the separation angle.

TABLE VIII. Variation of pressure drag with a slit width

| S/D ratio at slit inclination 0° | $C_{Dp}$ |
|---|---|
| 0 | 0.947 |
| 0.05 | 0.873 |
| 0.10 | 0.802 |
| 0.15 | 0.738 |
| 0.20 | 0.680 |
| 0.25 | 0.635 |

While analysis of an active control method, Delaunay and Kaiksis[3] calculated the global instability of the flow oscillation in terms of the maximum value of rms fluctuation intensities over the whole domain. The maximum values of rms fluctuation intensities (u′ and v′) are calculated by extracting the values in different locations of x/D as shown in Figure 17. Figure 17 reports the maximum fluctuation intensity of u′ and v′ values at Re=90 for the all S/D ratio keeping slit angle fixed at 0º. The present simulated results provide excellent predictions of $u'_{rms}$ and $v'_{rms}$, similar to the predictions reported by Delaunay and Kaiksis[3]. The small difference in quantitative values is due to the different blockage ratios of both cases. The interesting thing to notice here is that that the maximum values of fluctuation intensities weaken with an increase of the S/D ratio. Also, the decrease in the $u'_{rms}$ and $v'_{rms}$ stabilizes the flow, which corroborates the above discussion.

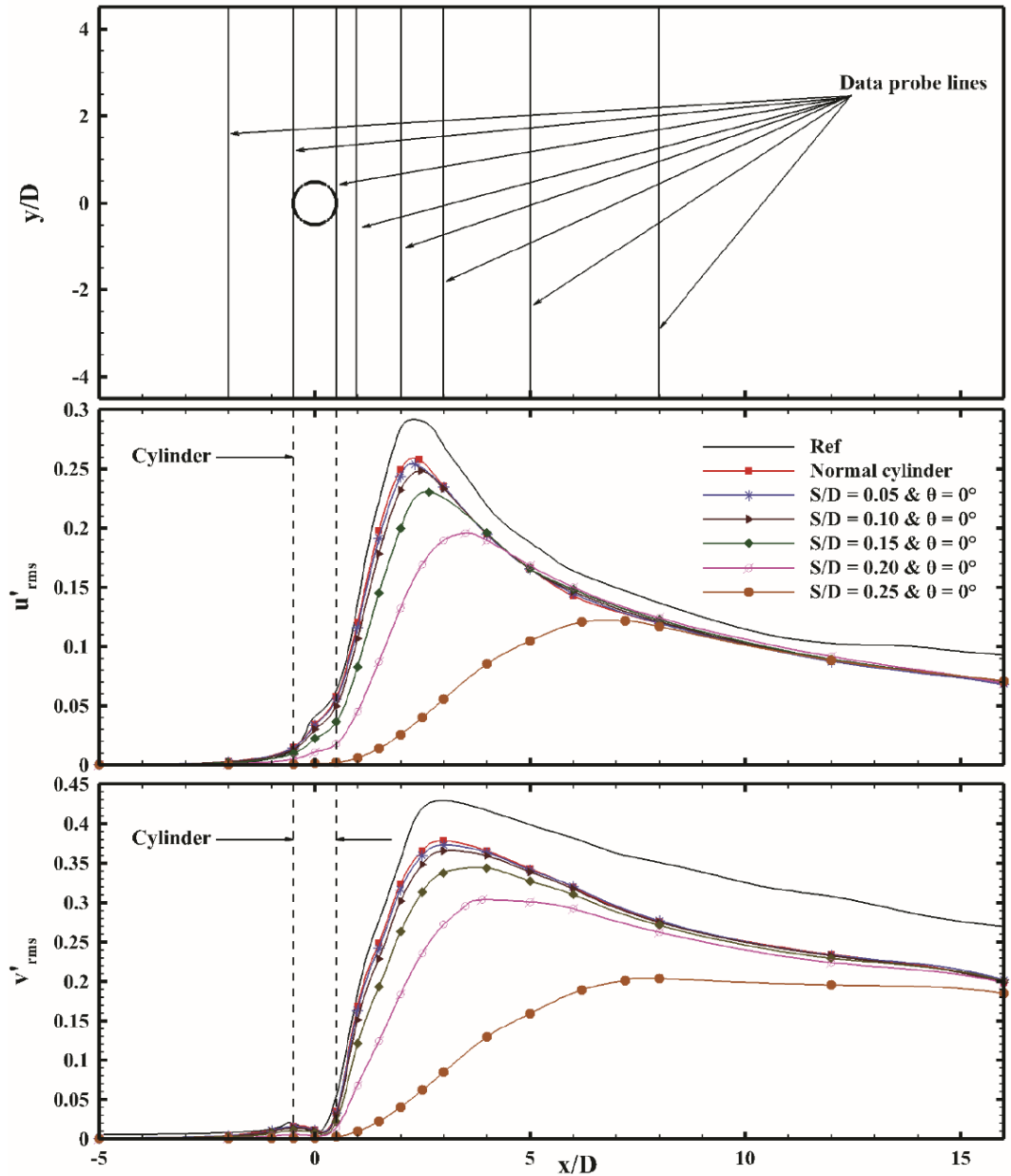

FIG. 17. The maximum rms fluctuation intensities of u′ and v′ along the line of constant x/D for various S/D. The reference data has been taken from Delaunay and Kaiksis[3].

## D. Slit angle variation with the incoming flow

The effect of slit angle with the incoming flow on critical Re for all S/D ratios is reported in this section. The simulations are performed for various slit ratios and varying slit angles 10º to 50º with an increment of step $\Delta\theta = 10º$. The variation of Re with slit inclination for a modified cylinder is depicted in Figure 18(a). The results indicate that the critical Re increases with angle up to S/D = 0.15, and surprisingly, it shows reverse behavior beyond this S/D ratio. S/D = 0.25 has steep decrement of critical Re with angle while S/D = 0.20 shows mild decrement. This phenomenon can be analyzed by investigating the separation angle on the cylinder and the fluctuation intensity of the flow with slit inclination. Figure 18(b) reports separation angle variation with Re for normal cylinder case, and the separation angle decreases with the Re. As noted and mentioned earlier, the increase of the suction pressure (or decrease in the separation angle) with Re develops the instability in the downstream of the cylinder and leads to unstable flow. So it can be concluded that the separation angle decrease with Re, which is more prone to unsteady flow. A similar trend has been depicted in slit cylinder cases. Figure 18(c) shows the separation angle variation with Re for the modified cylinder with S/D = 0.5 cases at slit inclination of $0^0$.

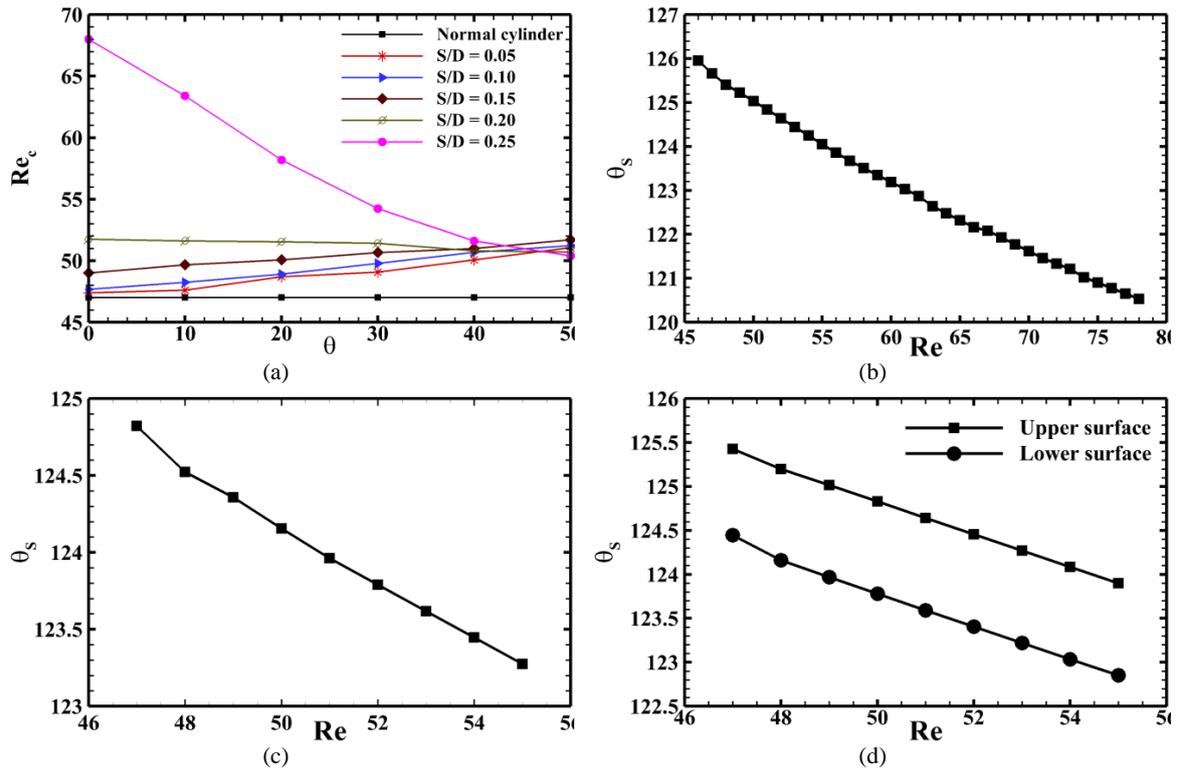

FIG. 18. (a) Variation of Re$_c$ with slit angle for different S/D (b) variation of separation with Re angle for unmodified cylinder (c) variation of separation angle for slit cylinder S/D = 0.05 & $\theta = 0^0$ (d) variation of separation angle S/D = 0.05 & $\theta = 10^0$

Figure 18(d) exhibits the separation angle variation on both the upper and lower part of the slit cylinder with S/D = 0.05 at the slit inclination of 10°. The upper surface has a higher value of the separation angle than the lower surface of the cylinder. The assessment yields that the flow around the lower surface is more inclined towards developing instability. And this instability can be controlled through the extra amount of energy provided by slot through the cylinder. The streamline patterns from Figure 19 indicate that flow through the slit is adding energy to the recirculation bubble on the lower surface of the cylinder for all the inclination of the slit angle for the S/D=0.05. Both the recirculation bubble has almost the same length for all the slit inclination for this case. The effect of slit inclination is negligible at recirculation bubble length for this case. A similar kind of flow physics has been found in the cases of S/D= 0.10 and 0.15 also. This is the primary reason behind the increase of the critical Re up to S/D=0.15.

To confirm the above observation, the maximum fluctuation intensity of $u'$ and $v'$ values are compared at Re=90 for the S/D = 0.05 ratio for different slit angle (Figure 20). The values $u'_{max}$ and $v'_{max}$ go down with the slit angle, which supports the above argument that the increase in slit angle may cause a decrease in the suction pressure and provides more stability to the flow over the cylinder. This is the reason behind the rise in the critical Reynolds number with a slit angle.

The critical Re reduces with slit inclination for the case S/D= 0.20 and 0.25. Figure 21 shows the streamline plots at various slit angles for S/D = 0.20, where both the recirculation bubbles have different size and strength for the 10° to 30°. This happens because the extra amount of energy through self-injecting jet goes one side (lower side of the cylinder) more than the other part of the cylinder, and this amount creates asymmetry in both the re-circulation bubbles. The asymmetry in the recirculation may develop the instability, and ultimately flow becomes unstable more rapidly in case of the slit inclination 10° to 30°. The slit inclinations 40° and 50° show symmetric recirculation bubbles because of the amount of extra energy is less in these cases as compared to the slit inclination 10° to 30°. So for this case, the value of critical Re is least, but it is still greater than the normal cylinder because of the self-injection of extra energy into the system. The instability calculation of the S/D = 0.20 for different slit angle is displayed in Figure 22 with the maximum fluctuation intensity of $u'$ and $v'$ values at Re=90. The observation suggests that the flow instability in the wake of the cylinder increases with the slit angle at this slit width ratio.

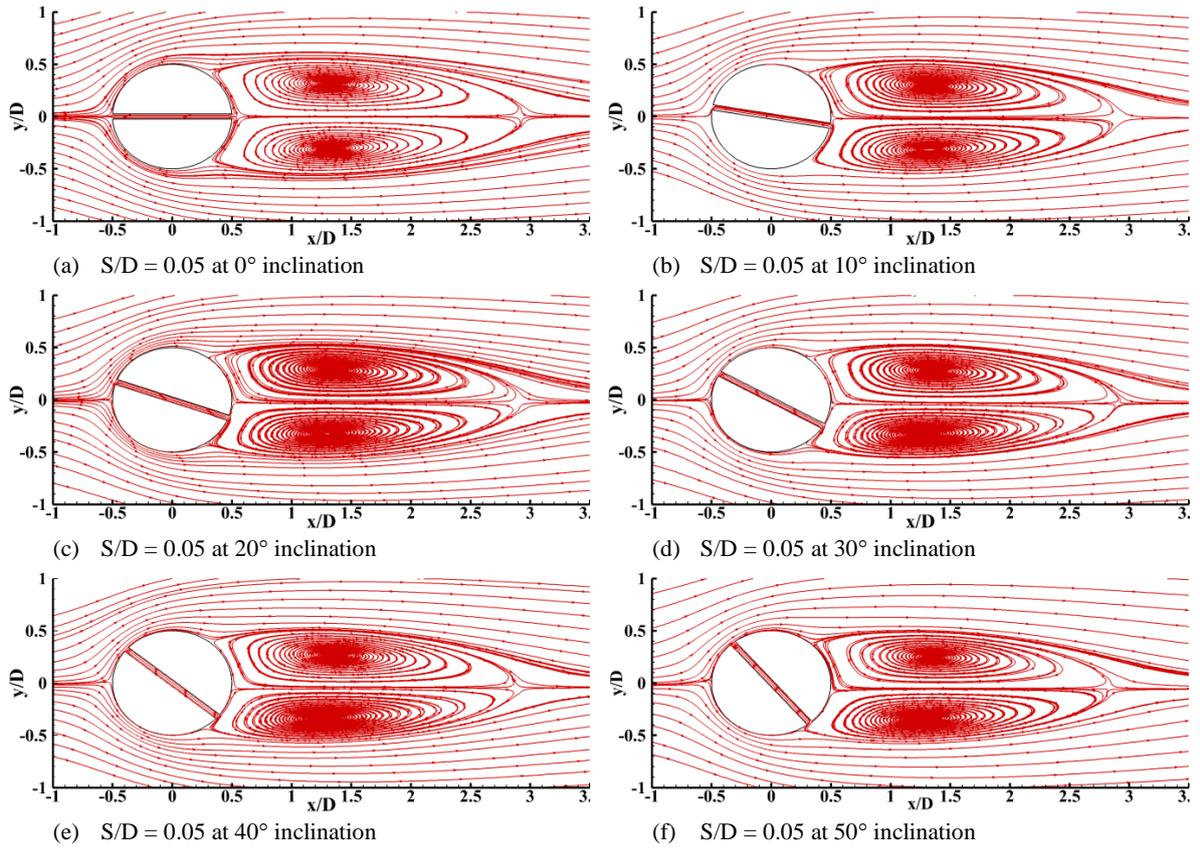

FIG. 19. Streamline pattern for S/D=0.05 with different slit inclination at Re = 47: (a) 0º, (b) 10º, (c) 20º, (d) 30º, (e) 40º, (f) 50º

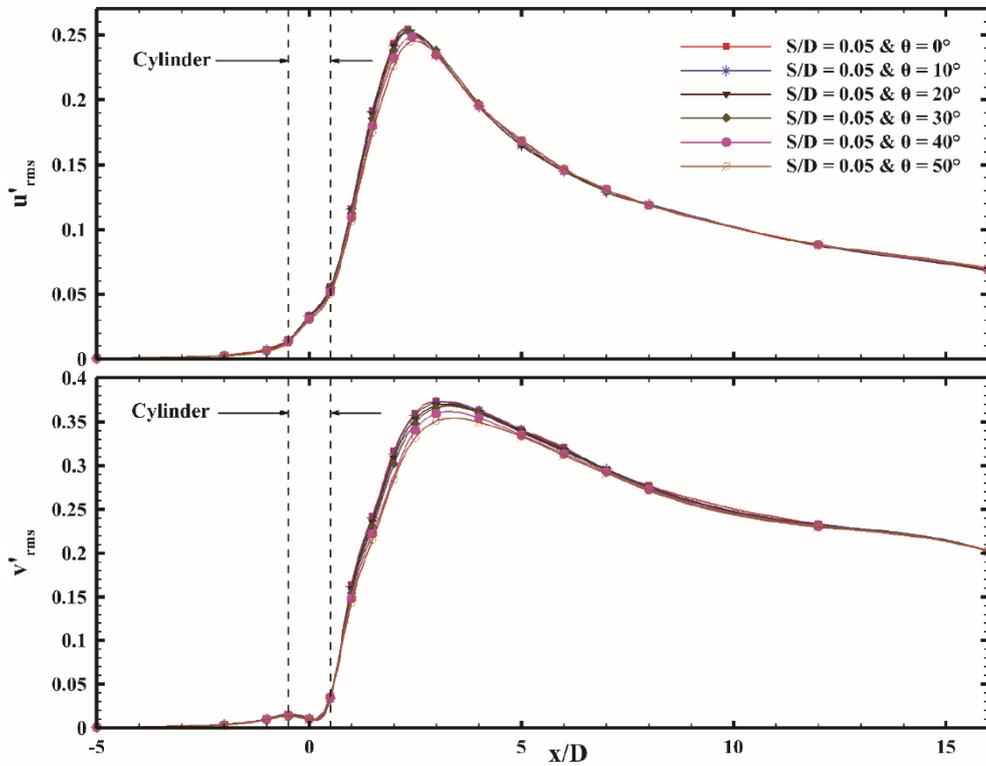

FIG. 20. The maximum rms fluctuation intensities of $u'$ and $v'$ along the line of constant x/D for various slit angle at S/D=0.05.

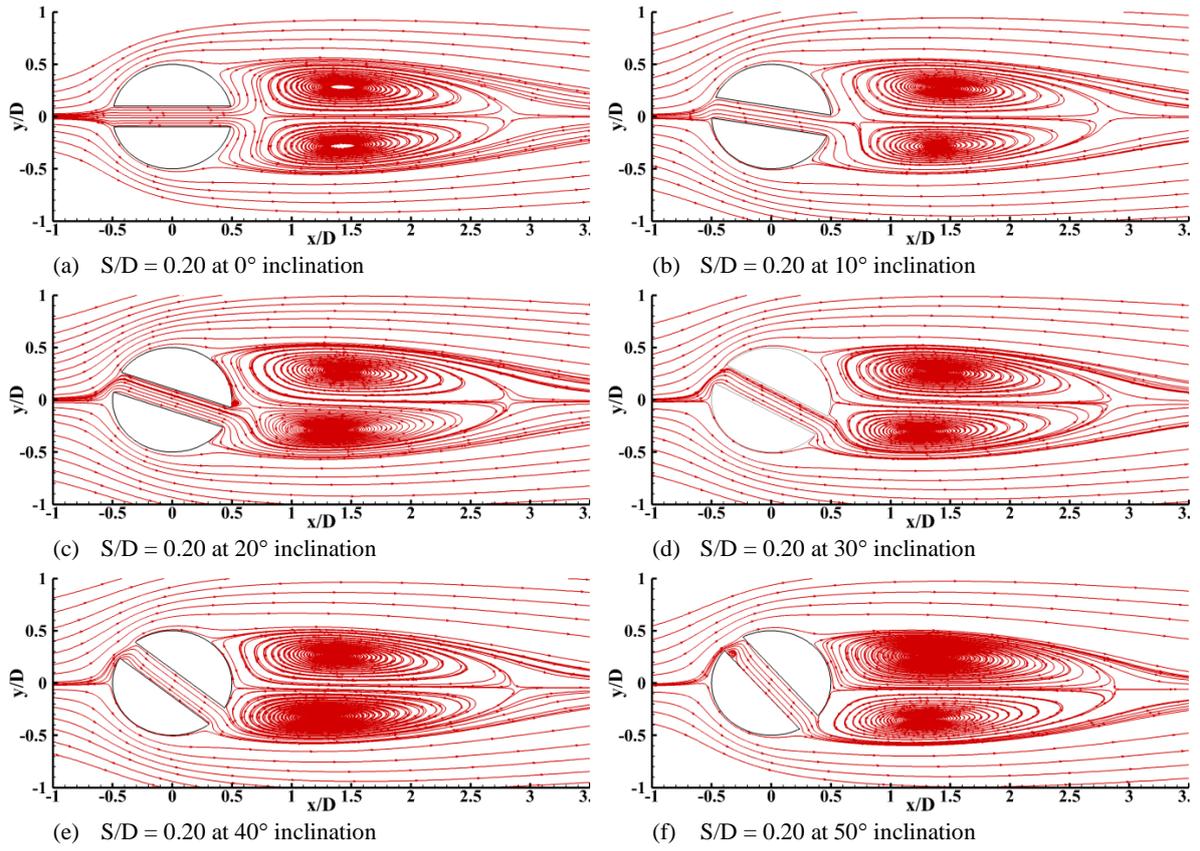

FIG. 21. Streamline pattern for S/D=0.20 with different slit inclination at Re = 47: (a) 0º, (b) 10º, (c) 20º, (d) 30º, (e) 40º, (f) 50º

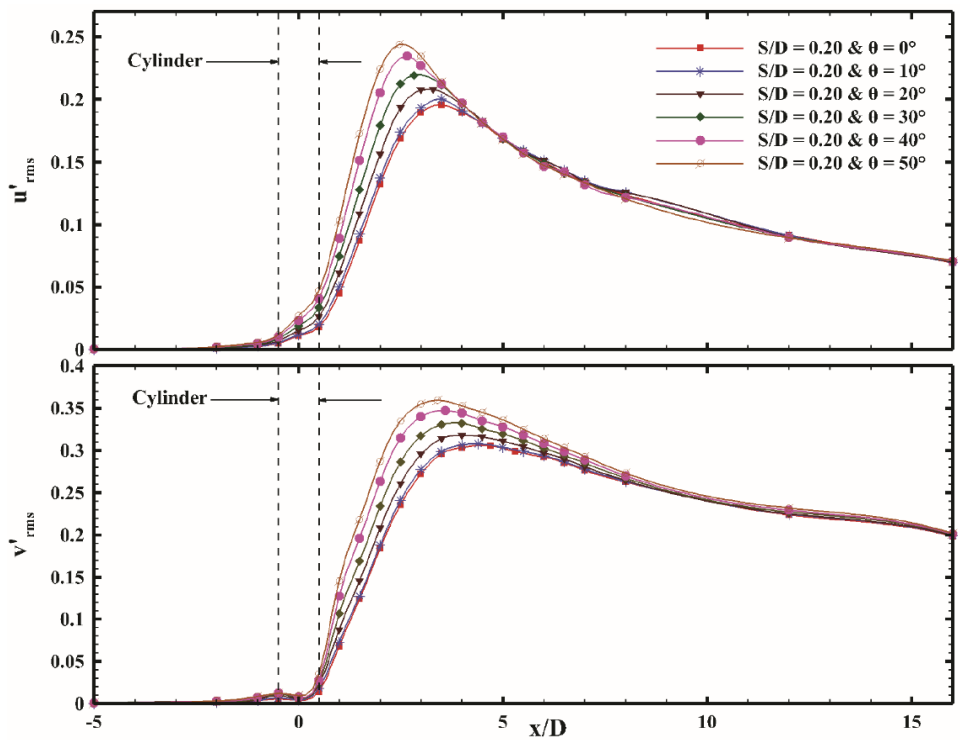

FIG. 22. The maximum rms fluctuation intensities of $u'$ and $v'$ along the line of constant x/D for various slit angle at S/D=0.20.

## E. Empirical Formula

A functional relationship has been developed between critical Reynolds number, slit width ratio and slit angle based on the simulated results of the modified and normal cylinder. The variation of critical Re with slit angle provides a different trend, as shown in section III(B). Figure 23 exhibits the relationship plot between the slit angle and the critical Reynolds number for S/D=0.05 and S/D=0.25. The observation yields linear progression fit between slit angle and $Re_c$ for various S/D ratios, which are represented in equations (3.7-3.12).

$$Re_c = 0 \times \theta + 47.02 \quad \text{for S/D} = 0 \tag{3.7}$$

$$Re_c = 0.0748\theta + 47.119 \quad \text{for S/D} = 0.05 \tag{3.8}$$

$$Re_c = 0.0744\theta + 47.56 \quad \text{for S/D} = 0.10 \tag{3.9}$$

$$Re_c = 0.0518\theta + 49.0533 \quad \text{for S/D} = 0.15 \tag{3.10}$$

$$Re_c = 0.02206\theta + 51.86 \quad \text{for S/D} = 0.20 \tag{3.11}$$

$$Re_c = 0.3555\theta + 66.89 \quad \text{for S/D} = 0.25 \tag{3.12}$$

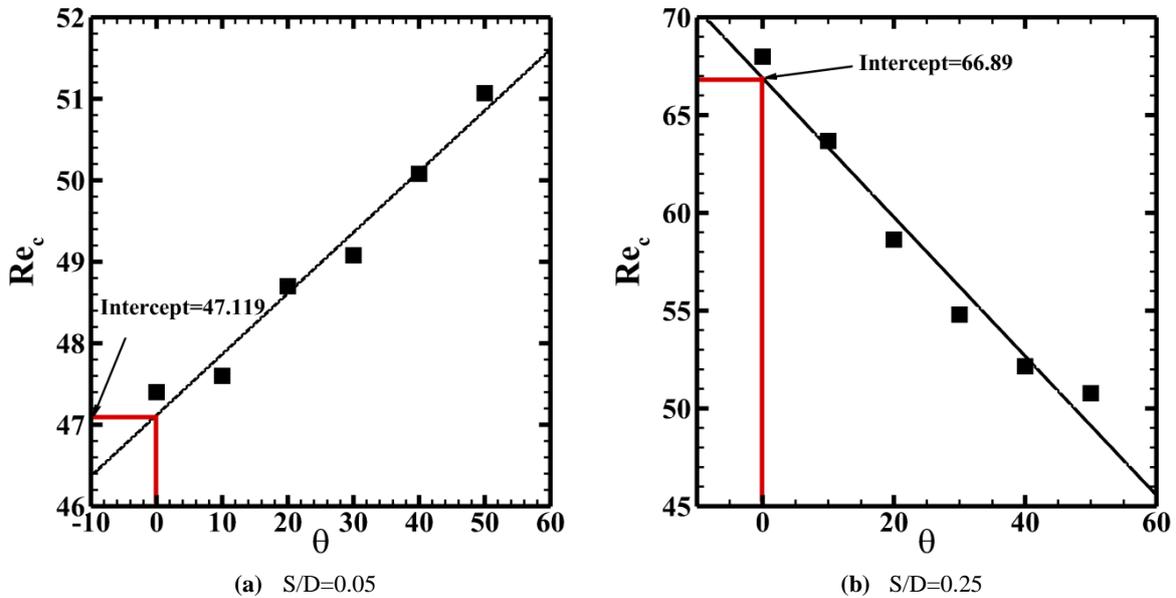

**(a)** S/D=0.05   **(b)** S/D=0.25

FIG. 23. Linear relationship curve between critical Reynolds numbers and slit angle for: (a)S/D=0.05, (b) S/D=0.25

Now, the slit width (S/D), slope, and intercept of linear variation are utilized to establish the general form of the empirical correlation. The slit width ratio (S/D) is first varied with the slope and then with the intercept of the linear relationship (critical Re with a slit angle) as displayed in Figure 24. It appears that a 4$^{th}$ order polynomial regression fit is found between S/D and the slope of the linear relations (Figure 24(a)), as depicted in equation (3.14). While Figure 24(b) confirms the quadratic relationship between S/D and intercept variation, as expressed in equation (3.15). Combining all of these, the general formula is represented in equations (3.13-3.15)

$$Re_c = f(S/D) \cdot \theta + g(S/D) \qquad (3.13)$$

$$f(S/D) = 3.524(S/D) - 55.507(S/D)^2 + 360.634(S/D)^3 - 870.797(S/D)^4 \qquad (3.14)$$

$$g(S/D) = 47.064 - 53.150(S/D) + 1538.778(S/D)^2 - 12535.2(S/D)^3 + 33986.33(S/D)^4 \qquad (3.15)$$

The developed empirical relation is verified against the different cases of simulated results, and the comparison between two are tabulated in Table IX. The values of critical Reynolds number through empirical correlation are closely matched with the simulated critical Reynolds number for all the cases. The maximum deviation in critical Reynolds number through empirical relation is observed around 3.288% (for S/D = 0.25 with slit angle $50^0$) only from simulated results.

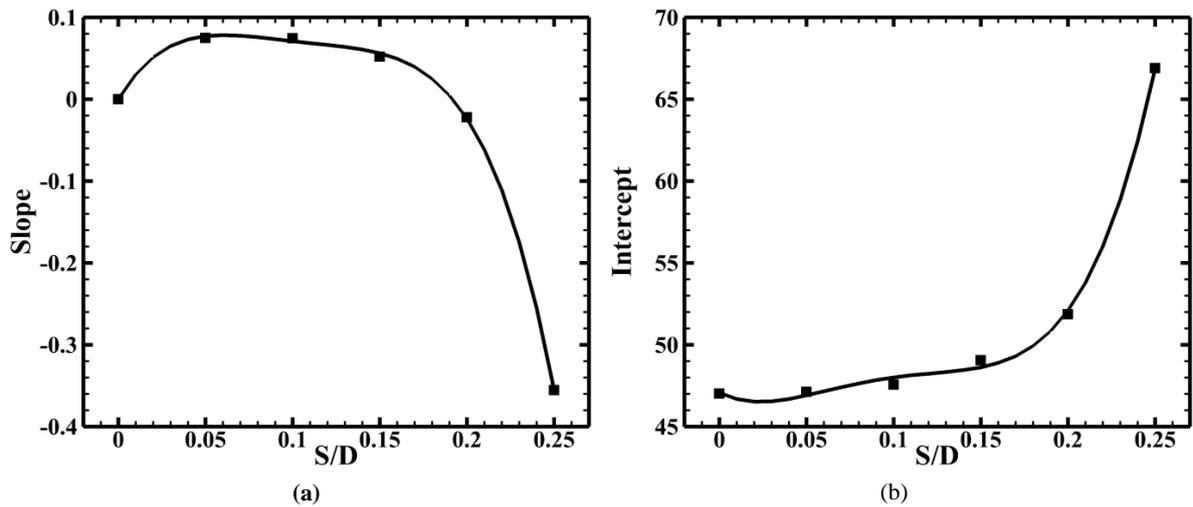

FIG. 24. (a) Variation of the slope with slit width ratio, S/D, (b) Variation of intercept with slit width ratio

TABLE IX. Comparison of critical Reynolds number evaluated from empirical formula and SAA

| S/D Ratio | Slit Angle ($\theta$) | $Re_c$ from an empirical formula | $Re_c$ from computation | % error |
|---|---|---|---|---|
| 0.05 | 10 | 47.6698 | 47.6 | -0.14664 |
| 0.10 | 20 | 49.41858 | 48.91 | -1.03982 |
| 0.15 | 30 | 50.29464 | 50.67 | 0.740785 |
| 0.20 | 40 | 51.13734 | 50.8 | -0.66406 |
| 0.25 | 50 | 49.11008 | 50.78 | 3.288541 |

## IV. Conclusion

The present study investigates the critical Reynolds number for a passive control case with a slit through the cylinder for varying slit ratios and slit angles. First, the validation of the flow field over a normal cylinder is carried out at Re =100, where simulations provide a good prediction of the flow field as compared to the previous studies. The Hopf bifurcation is found to be at Re= 47 for an unmodified cylinder using flow visualization method (FVM) and saturation amplitude analysis (SAA). The critical Reynolds number for normal cylinder also is also evaluated using Stuart-Landau equation and global stability analysis, which provides an excellent match with FVM and SAA. The cylinder has been modified with a slit along with the incoming flow. The passive control significantly increases the $Re_c$. At slit angle = $0^0$, critical Re is directly proportional to the slit ratio S/D. The results find the highest critical Reynolds number (Re=68) for S/D=0.25 at the slit inclination of $0^0$. The additional energy through the slit reduces the base suction of the cylinder. It controls the instability developed in the wake of the cylinder. The separation angle on the cylinder also increases with the slit width. Results of variation of slit inclination can be divided into two regions. From S/D=0.05 to S/D= 0.15, the global instability decreases with slit inclination; as a result, the critical Re increases with slit angle. For the range S/D=0.15 to 0.25, the extra amount of energy at an angle creates instability in the wake of the cylinder, which consequently reduces the critical Reynolds number.

**ACKNOWLEDGMENTS**

Simulations are carried out on the computers provided by the Indian Institute of Technology Kanpur (IITK) (www.iitk.ac.in/cc), and the manuscript preparation, as well as data analysis, has been carried out using the resources available at IITK. This support is gratefully acknowledged.